# FULL SCALE DYNAMIC RESPONSE OF A RC BUILDING UNDER WEAK SEISMIC MOTIONS USING EARTHQUAKE RECORDINGS, AMBIENT VIBRATIONS AND MODELLING

## Clotaire MICHEL[1], Philippe GUEGUEN[2]
## Saber EL AREM[3], Jacky MAZARS[3], Panagiotis KOTRONIS[3]

[1] Applied Computing and Mechanics Laboratory (IMAC), Ecole Polytechnique Fédérale de Lausanne, Switzerland

[2] LGIT, University Joseph Fourier, CNRS, LCPC, France

[3] 3S-R, University Joseph Fourier, CNRS, INPG, France

Laboratoire de Géophysique Interne et Tectonophysique (LGIT)

Université de Grenoble

1381 rue de la Piscine

38041 Grenoble Cedex 9 FRANCE

Corresponding author
P. Guéguen
LGIT
BP 53
38041 Grenoble cedex 9
France
pgueg@obs.ujf-grenoble.fr

**Abstract**

In countries with a moderate seismic hazard, the classical methods developed for strong motion prone countries to estimate the seismic behaviour and subsequent vulnerability of existing buildings are often inadequate and not financially realistic. The main goals of this paper are to show how the modal analysis can contribute to the understanding of the seismic building response and the good relevancy of a modal model based on ambient vibrations for estimating the structural deformation under moderate earthquakes. We describe the application of an enhanced modal analysis technique (Frequency Domain Decomposition) to process ambient vibration recordings taken at the Grenoble City Hall building (France). The frequencies of ambient vibrations are compared with those of weak earthquakes recorded by the French permanent accelerometric network (RAP) that was installed to monitor the building. The frequency variations of the building under moderate earthquakes are shown to be slight (~2%) and therefore ambient vibration frequencies are relevant over the elastic domain of the building. The modal parameters extracted from ambient vibrations are then used to determine the 1D lumped-mass model in order to reproduce the inter-storey drift under weak earthquakes and to fix a 3D numerical model that could be used for strong earthquakes. The correlation coefficients between data and synthetic motion are close to 80% and 90% in horizontal directions, for the 1D and 3D modelling, respectively.

# 1. Introduction

Since Omori [1], certainly the first having recorded ambient vibrations in Japanese buildings for earthquake engineering applications, an abundant scientific literature has been published showing the interest, advantages but also limitations in performing such weak motion recordings in buildings (e.g., [2]; [3]; [4]; [5]). Most of the studies were focused on the determination of the resonance frequencies for earthquake and mechanical engineering. Crawford and Ward [4] and Trifunac [5] showed that ambient vibration-based techniques were as accurate as active methods for determining vibration modes and much easier to implement for a large set of buildings. More recently, Hans et al. [6] showed that the vibration modes extracted from ambient vibrations and active methods were quite similar in the $10^{-5}$ to $10^{-2}$g range of loading.

Nevertheless, it is well known that under strong seismic loading, the resonance frequency of existing buildings decreases, thus modifying the seismic demand that depends on the period of the building. Celebi [7] and Irie et al. [8] observed this permanent decrease of the resonance frequency computed using strong seismic motion recorded in Californian buildings compared to the ambient vibrations. For strongest motions, a recent scientific literature describes the permanent decrease of the structural frequency of buildings due to structural damage (e.g. [9]; [10]; [11]) as well as the transient drop of the frequency due to the closing-opening process under shaking of pre-existing cracks [12].

In the last 20 years, modal analysis techniques in civil engineering applications have been considerably improved thanks to technical (instrumentation, computers) and theoretical developments in the electrical and mechanical engineering fields ([13]; [14]; [15]; [16]). The frequencies at which vibration naturally occurs, and the modal shapes which the vibrating system assumes, are properties of the system. They can be determined analytically using Modal Analysis. The analysis of vibration modes is a critical component of the design, for understanding the behaviour of complex structures and fixing their elastic properties by means of their modal parameters (frequency, damping and modal shape). These are also the main parameters controlling seismic building response and vulnerability since "the natural period of vibration is the single most informative fact about the internal structure of a building. Two structures with the same mass distribution and the same fundamental period may experience shear forces of appreciably different magnitudes if the internal structures (mode shapes) are different" (after [3]). The major difficulty in the dynamic response assessment of existing buildings is the lack of available data such as quality of the materials, structural plans, ageing and structural integrity. In such cases, the classical tools in earthquake engineering may turn out to be very expensive as for countries with a moderate seismic hazard, like France. In such areas, the cost of enhanced methods is not justified, let alone for assessments on large sets of buildings, even though the hazard described in the seismic design codes would be a motive for it. Models based on the experimental modal values, from the simplest analytical model to the most comprehensive finite-element model, can be used in evaluating the deformation that occurs in buildings during moderate earthquakes. These simulations can be the linear starting point of a more extensive analysis of the non-linear response for seismic vulnerability assessment (e.g., [17]).

The main goals of this paper are to show how the experimental modal analysis can contribute to the understanding of the seismic building behaviour and the good relevancy of a modal model based on ambient vibrations for estimating the structural deformation under moderate earthquakes. More specifically, we study the response of the Grenoble City Hall (France), a 13-storey reinforced concrete building, using ambient vibration tests and the network of permanent accelerometric monitoring stations installed by the French Permanent Accelerometric Network. After briefly describing the structural design of the building and the experimental networks used (ambient vibration survey and accelerometric network), the results of the modal analysis of the building using weak earthquakes recorded in the structure are compared with those of the ambient vibration survey. The accelerometric data observed at the top of the building are then compared to those predicted using a 1D lumped-mass model adjusted using the modal analysis results obtained from ambient vibration recordings, and a 3D numerical model based on multifiber beam elements.

## 2. The Grenoble City Hall building

The city of Grenoble is located in the northern French Alps (Fig. 1), one of the most seismic-prone areas in France ($a_N$=1.5 m/s$^2$ for the old national seismic design code PS92, and $a_g$=1.6m/s$^2$ for the French national annexes of the EC8 code). Several strong historical events have occurred in the surrounding area and the regional seismic network (Sismalp, http://sismalp.obs.ujf-grenoble.fr) indicates an active fault along the Belledonne range, 15 km from the city [18]. Furthermore, the city is founded on a very deep sedimentary basin giving rise to strong site effects ([19]; [20]). The moderate seismicity and sedimentary contrast, coupled with the high exposure due to the number of inhabitants, hi-tech and nuclear facilities, make Grenoble a national case study for seismic risk analysis.

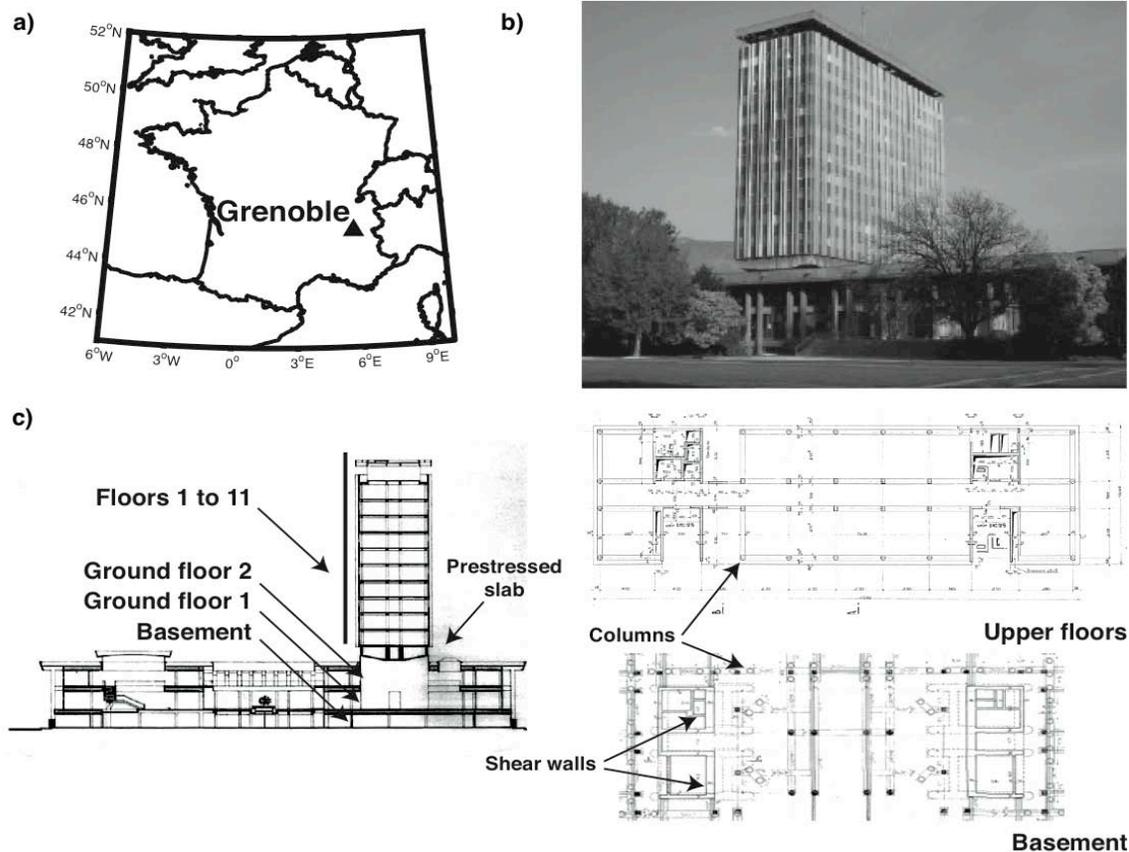

**Figure 1:** a) Location of Grenoble in France b) Grenoble City Hall, viewed from the southeast; c) Cross section and plan view of the basement storey and of a current storey of the tower.

The Grenoble City Hall is a reinforced concrete (RC) structure completed in 1967 (Fig. 1). It is divided into two parts: a 2-story horizontal building and an independent 13-story tower that is the object of this study. No structural connections are observed between the main tower and secondary small building. The tower has a 44 m by 13 m (L, T respectively) plan section and rises 52 m above the ground. The inter-storey height is regular between the $3^{rd}$ and $12^{th}$ floors (3.2 m) and higher for the $1^{st}$ (4.68 m) and $2^{nd}$ storey (8 m), above which there is a prestressed slab of 23 m span supported by two inner cores. These cores, consisting of RC shear walls, enclose the stair wells and lift shafts and are located at two opposite sides of the building. The structural strength system combines these shear walls with RC frames with longitudinal beams bearing the full RC floors. The glass frontage is fastened to a light steel framework placed on the external perimeter of the structure. The foundation system consists of deep piles, anchored in an underlying stiff layer of sand and gravel. This structure did not benefit from an earthquake design and the design report was not available. The design of this structure did not include seismic forces and any information on yielding story drift or yielding base shear coefficient was available.

## 3. Weak seismic motion and ambient vibration surveys

Since November 2004, the building has been monitored by six accelerometric stations, three on the ground floor called OGH1, OGH2 and OGH3 and three on the $13^{th}$ floor called OGH4, OGH5 and OGH6 (Fig. 2). This instrumentation is part of the French Permanent Accelerometric Network (RAP), which is in charge of recording, collecting and disseminating accelerometric data in France [21]. The City Hall building array is managed by the Geophysical Laboratory (LGIT) of Université Joseph Fourier (Grenoble, France). Each station consists of one 3C Episensor (Kinemetrics) accelerometer connected to a MiniTitan 24-bit digital acquisition system (Agecodagis). The horizontal components are oriented along the longitudinal and transverse directions of the building, with the longitudinal direction having an azimuth of 327°N. The sampling rate is 125 Hz and the recordings are divided into files of 2 minutes in length. Time is controlled by a GPS receiver located on top of the building. The stations are connected via an Ethernet hub allowing data transfer from each station to the computer located in the basement of the building and time synchronization. This computer is permanently online for remote data control and station management. The dial-up data retrieval system at the LGIT extracts the data from the continuous recordings in accordance with a list of epicentres provided by the French national seismological survey (RéNaSS). All the data are integrated in the online database of the RAP and can be retrieved in ASCII, SAC and SEED format (http://www-rap.obs.ujf-grenoble.fr). Within the context of this study, a temporary experiment was also performed for determining the full-scale behaviour of the structure under ambient vibrations. A Cityshark II station [22] was used for the simultaneous recording of 18 channels. Six Lennartz 3D 5s velocimeters were used for this purpose, having a flat response between 0.2 and 50 Hz. Eight datasets were recorded, corresponding to 36 different points in the building, i.e., at least two points per floor. One sensor was installed on top of the building to serve as reference instrument for all the datasets. This reference point is necessary in order to normalize and combine all the components of the modal shape [23]. The first frequency was roughly estimated to be close to 1 Hz, so a 15 min recording time was selected for each set, corresponding to more than 1000 periods, at a 200 Hz sampling rate.

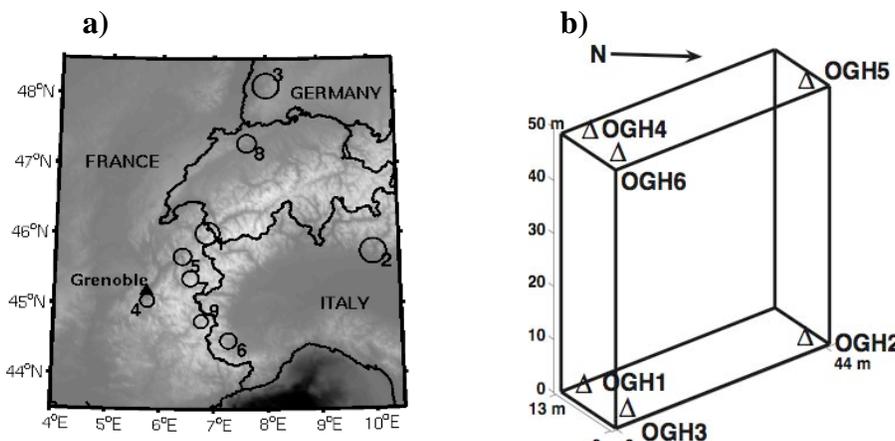

**Figure 2.** a) Location of the earthquake epicenters used in this study and located by the French national seismological survey (RéNaSS). b) Location of the accelerometers of the French Accelerometric Network (RAP) in the City Hall.

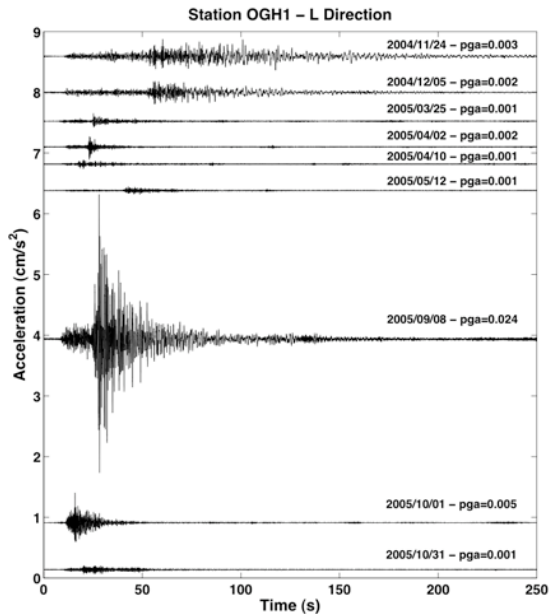
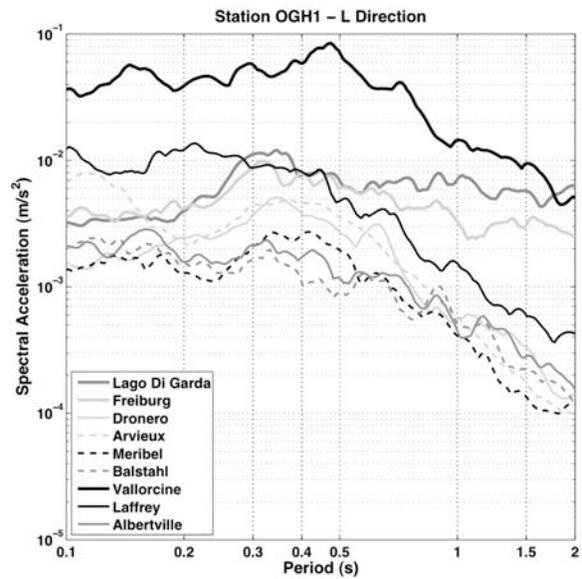
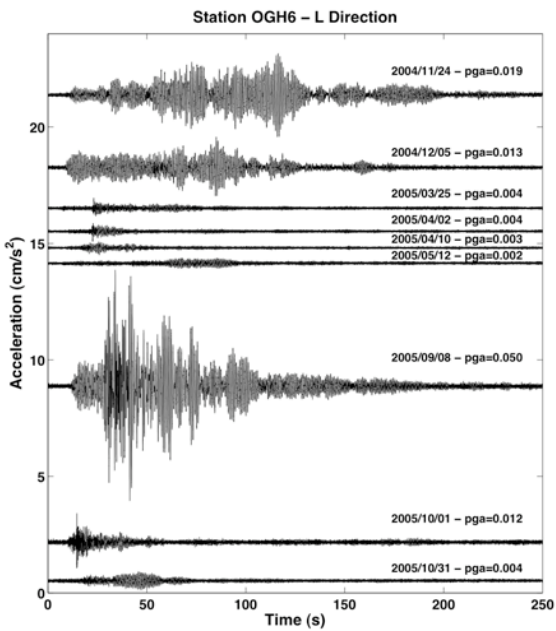

**Figure 3.** Examples of accelerometric time history (left) of the nine earthquakes recorded in Grenoble City Hall at the OGH6 roof station (lower row) and at the OGH1 (upper row) ground station in the longitudinal L directions. All waveforms are plotted in relative mode and scaled by the maximum amplitude of each station/component pair. Corresponding response spectra (right).

## 4. Earthquake recordings

Since 2004, more than 25 earthquakes have been recorded by the permanent building array. Nine earthquakes were selected (Fig. 2a) with a signal to noise ratio greater than 3 in the 0.6-5 Hz frequency band corresponding to a PGA from 0.6 to 23 mm/s² (Fig. 3). The French national seismological survey (RéNaSS) located these earthquakes in the most active zones of the western part of the Alps, corresponding to the Northern (events #1, 4, 5, 7) and Southern (events #6, 9)

French Alps, the Italian Alps (event #2) and the Rhine Graben (events #3, 8). Event #4 is located on the Belledonne Border Fault system [18]. Table 1 summarises the Horizontal Peak Ground Accelerations (PGA), Velocities (PGV) and the Arias Intensity ($I_{ag}$), i.e. the energy of the accelerogram a(t), defined as follows [24]:

$$I_{ag} = \frac{\pi}{2g} \int_0^\infty [a(t)]^2 dt \qquad (1)$$

**Table 1**. Characteristics of the earthquakes and their recordings in Grenoble City Hall used in this study (*T,t* indexes correspond to the top of the building; *G,g* indexes correspond to the ground)

| # | Event characteristics (RéNaSS) | | | | | Epicentral Distance (km) | Maximum Acceleration (mm/s²) | | Maximum Velocity (mm/s) | | Maximum Arias Intensity (μm/s) | | Max. Drift (10⁻⁶) |
|---|---|---|---|---|---|---|---|---|---|---|---|---|---|
| | Location | Long. | Lat. | $M_L$ | Date | | PGA | PTA | PGV | PTV | $I_{ag}$ | $I_{at}$ | $D_m$ |
| 1 | Vallorcine | 6.87 | 46.01 | 4.9 | 2005/09/08 | 127.3 | 22.94 | 107.42 | 1.376 | 11.80 | 106.71 | 7943.1 | 30.59 |
| 2 | Lago di Garda (Italy) | 10.01 | 45.74 | 5.5 | 2004/11/24 | 339.8 | 3.10 | 22.89 | 0.695 | 3.10 | 9.39 | 1229.3 | 8.30 |
| 3 | Freiburg (Germany) | 8.00 | 48.11 | 5.3 | 2004/12/05 | 368.3 | 1.88 | 19.17 | 0.294 | 2.21 | 3.46 | 301.1 | 5.74 |
| 4 | Laffrey | 5.75 | 45.05 | 3.1 | 2005/10/01 | 15.2 | 4.70 | 11.62 | 0.159 | 0.94 | 2.08 | 50.2 | 2.13 |
| 5 | Albertville | 6.40 | 45.68 | 3.6 | 2005/10/31 | 75.6 | 0.78 | 3.73 | 0.053 | 0.53 | 0.18 | 17.4 | 1.40 |
| 6 | Dronero (Italy) | 7.27 | 44.48 | 3.5 | 2005/03/25 | 144.5 | 1.31 | 4.70 | 0.084 | 0.52 | 0.33 | 17.5 | 1.38 |
| 7 | Meribel | 6.56 | 45.36 | 3.4 | 2005/04/10 | 67.5 | 0.88 | 5.10 | 0.040 | 0.56 | 0.14 | 18.1 | 1.25 |
| 8 | Balstahl (Switzerland) | 7.63 | 47.29 | 3.9 | 2005/05/12 | 275.7 | 0.56 | 3.73 | 0.033 | 0.47 | 0.13 | 20.6 | 1.21 |
| 9 | Arvieux | 6.76 | 44.75 | 3.1 | 2005/04/02 | 94.2 | 1.98 | 3.47 | 0.084 | 0.35 | 0.31 | 9.2 | 0.85 |

This table also gives a set of parameters computed for describing the building motion: the Peak Top Accelerations (PTA), Velocities (PTV), the Arias Intensity at the top ($I_{at}$) and the maximum drift ($D_m$) between the top and the ground floor. This last parameter, calculated as the difference between top and base displacements divided by the building height, is used by many methods as threshold criteria for damage assessment (e.g. [25]). The displacement is obtained from the acceleration by a double integration of the filtered signal between 0.1 and 40 Hz (Butterworth filter of order 4). The values of parameters on the ground and at the top displayed in Tab. 1 correspond to the maximum values given by the three stations on the ground floor and on top of the building, respectively (unfiltered data).

The earthquakes are sorted with respect to $D_m$ in decreasing order. Their values range from $10^{-6}$ to $3\;10^{-5}$. They are low compared to the limit tensile strain of concrete (around $10^{-4}$) which means that the building can be considered to be undamaged by the 9 earthquake analysed. The peak acceleration and velocity at the top are 1.8 to 10.2 times and 4.1 to 14.2 times the peak acceleration and velocity at the base, respectively.

One remarkable event is the Vallorcine (Haute-Savoie, France) $M_L$=4.9, September 8[th] 2005 earthquake (#1), which is the strongest event recorded in the Grenoble City Hall since monitoring started. Only minor damage and rock falls occurred in the epicentre zone, but it was strongly felt in the Alps and especially in the Grenoble basin, due to the strong site effects ([19]; [20]). Although no damage was observed in Grenoble more than 120 km from the epicentre, people working above the third level (i.e. above the prestressed slab) spontaneously evacuated the City Hall. However, the drift observed for this event (Tab. 1) is 3 10$^{-5}$, i.e. three times lower than the concrete limit tensile strain. Guéguen and Bard [26] distinguish four independent structural modes of deformation: structural drift, torsion, base rocking and relative motion of the foundation. The contribution of each has been analysed in the time domain for both horizontal directions, except the relative motion of the foundation due to the lack of free field sensors. In this paper, we display hereafter only the analysis in the L-direction for the Vallorcine (Event #1) earthquake (Fig. 4).

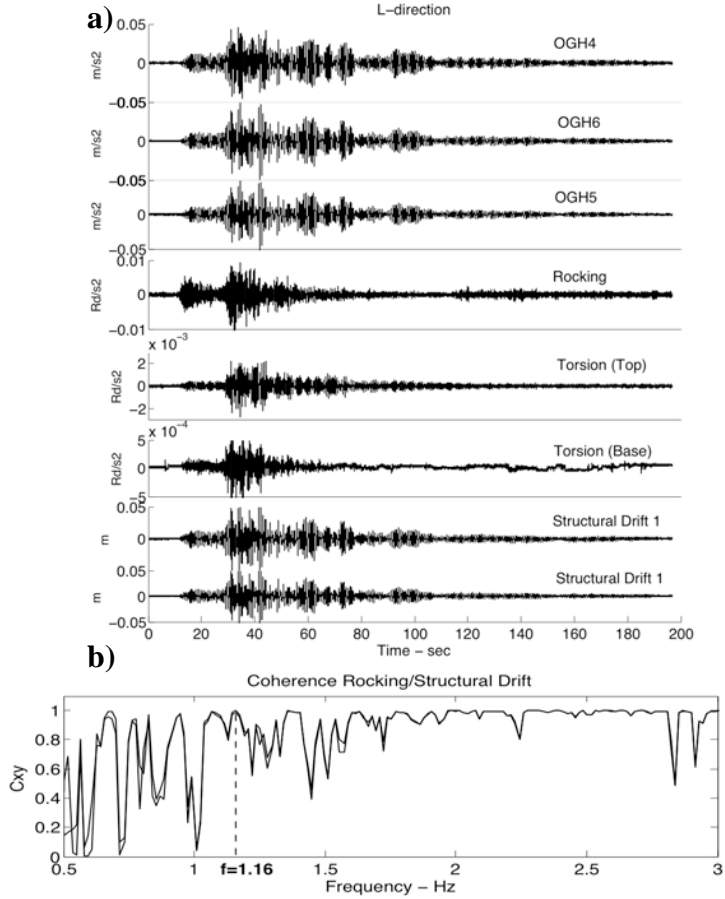

**Figure 4.** a) Time histories of horizontal displacement at roof level station (OGH4, OGH6 and OGH5) in the L-direction and comparison with the components of the main structural deformation recorded during the Vallorcine (Haute-Savoie, France) $M_L$=4.9, September 8$^{th}$ 2005 earthquake (see text for comprehension); b) Coherence Crd between rocking and structural drift computed in the L-directions for the Vallorcine earthquake using the two sets of station used to compute the structural drift (thin line: structural drift 1; thick line: structural drift 2).

Torsion at the top and at the base are derived from the difference between L-components recorded at each corner and normalized by the distance L between two corners (i.e. (OGH6$^L$-OGH5$^L$)/L and (OGH3$^L$-OGH2$^L$)/L, respectively). Torsion at the basement (Fig. 4) is very insignificant relative to the horizontal and rocking acceleration (about 100 times less in the L-direction), as previously observed by Meli et al. [27] and Guéguen and Bard [26] in full-scale buildings. Torsion at the top for this building is also low compared to horizontal drift and rocking. As mentioned in Bard [28] and Guéguen and Bard [26], assuming a rigid behaviour of the foundation rocking acceleration R$^L$ is computed as follows:

$$R^L = H \frac{OGH3^Z - OGH2^Z}{L} \qquad (2)$$

This rocking corresponds to the total rocking of the basement, including the rocking of the soil, which is in general considered negligible. The maximum value of rocking (Fig. 4) corresponds to about 10 mm/s$^2$, namely around 20% of the peak top acceleration of the structure (Tab. 1). As reported in various papers (e.g. [28]; [27]), buildings founded on soft soils usually exhibit significant rocking owning to soil-structure interactions (SSI). This is true even for structures founded on deep foundations.

The structural drift D$^L$, which corresponds to the fixed-base structure behaviour, is computed by subtracting the total acceleration of the foundation (i.e. rocking plus horizontal acceleration including the input acceleration) from the building top acceleration (i.e. D1$^L$=OGH6$^L$ - OGH3$^L$ - R$^L$ and D2$^L$=OGH5$^L$ - OGH2$^L$ - R$^L$). The maximum of the structural drift (Fig. 4) represents about 100% of the peak acceleration recorded at the top of the structure. Note that, whichever set of stations considered in the same direction, the time history of the structural drift is quite similar.

The coherence $C_{rd}$ between rocking $R^L$ and structural drift $D^L$ is plotted in Fig. 4b. It is computed as follows:

$$C_{rd} = \frac{\left|S_{xy}^2\right|}{S_{xx}S_{yy}} \qquad (3)$$

where $S_{xx}$ and $S_{yy}$ are the power spectral density (PSD) of the signals x and y respectively, and $S_{xy}$ is the cross-PSD of x and y. In our case, x corresponds to the rocking motion and y to the structural drift. For the Vallorcine earthquake, high coherence is observed at the frequencies of the structure, which will be detailed in the following, between rocking and structural drift (more than 95%) in L-directions, which attests SSI effects [28].

Finally, this analysis showed that the building motion was mainly following its horizontal drift vibration modes but with a non-negligible effect of SSI (rocking), which reaches 20% of the top acceleration in the case of Vallorcine earthquake. SSI effects were also found using the lowest earthquake data. This SSI is associated to the presence of a very soft soil in the 20 first meters depth, made of very soft clay characterized by S-wave velocity close to 200 m/s [20].

## 5. Ambient vibration recordings processing using Frequency Domain Decomposition

The experimental behaviour of buildings can be formalized in a complete and detailed model using modal analysis of ambient vibrations (called Operational Modal Analysis, OMA). OMA is now widely used in civil engineering applications (e.g. [14]; [16]) to understand the linear behaviour of structures in terms of vibration modes. In order to extract the modal parameters of the structure from ambient vibration recordings, the Frequency Domain Decomposition (FDD) method [29] was used in this paper. This method is able to decompose modes, even if they are very close. The first step of this method is to calculate the PSD matrices for each dataset. The Welch method [30] was used for this purpose, for which Fourier Transforms of the correlation matrices on overlapping Hamming windows are averaged over the recordings. The length of the time windows has been increasingly tested until a value of $2^{15}$ samples giving a frequency precision of $200/2^{15}=0.006$ Hz. Given that 18 channels are recorded simultaneously, the size of these matrices is 18x18 for each frequency. Only a limited number of modes (frequencies $\lambda_k$, mode shape vectors $\{\Phi_k\}$) have energy at one particular angular frequency $\omega$ noted Sub($\omega$). It can be shown [29] that the PSD matrices of the sensors $[Y](\omega)$ using the pole/residue decomposition take the following form:

$$[Y](\omega) = \sum_{k \in Sub(\omega)} \frac{d_k \{\Phi_k\}\{\Phi_k\}^T}{j\omega - \lambda_k} + \frac{\bar{d}_k \{\bar{\Phi}_k\}\{\bar{\Phi}_k\}^T}{j\omega - \bar{\lambda}_k} \qquad (4)$$

with $d_k$ a constant and $j^2 = -1$. Moreover, a singular value decomposition of the estimated PSD matrices at each frequency can be performed:

$$[\hat{Y}](\omega) = [U_i][S_i][\bar{U}_i]^T \qquad (5)$$

Identification of Eq. 4 and 5 shows that the modulus of the first singular value gives a peak for an ω value corresponding to a resonance frequency $\omega_k$ linked to the continuous-time eigenvalues $\lambda_k = -\xi_k \omega_k \pm j\omega_k \sqrt{(1-\xi_k^2)}$ (Fig. 5a). Furthermore, if Sub(ω) has only one or two geometrically orthogonal elements, the first two singular vectors are proportional to the modal shapes.

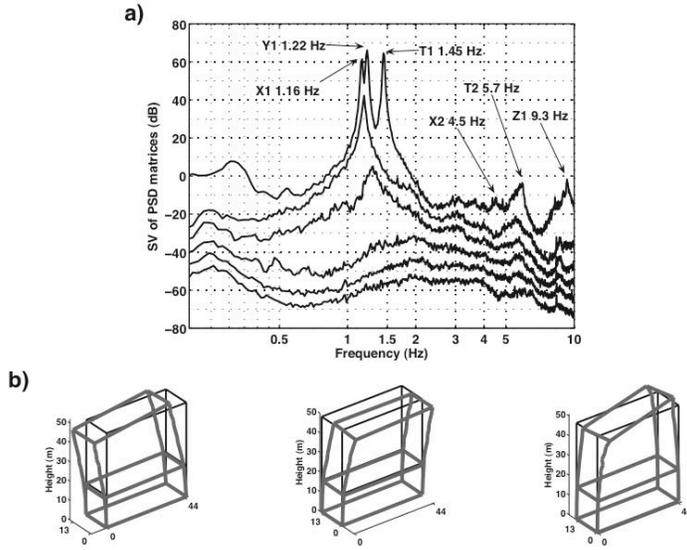

**Figure 5.** a) Spectrum (mean value of the 8 datasets of the first 6 singular values of the PSD matrices) of the structure under ambient vibrations computed using Frequency Domain Decomposition (FDD) (Brincker et al., 2001a). b) first 3 structural modes of the structure obtained using FDD (from left to right: longitudinal bending, transverse bending and torsion).

In practice, buildings are often equally stiff in both longitudinal and transverse directions so that the first modes in each direction are very close each other. The FDD method is capable of decomposing these modes contrary to the traditional "Peak Picking" method. Moreover, this method can be enhanced [31] to select the complete mode "bell", and consequently its damping ratio, by comparing the mode shape at the peak to the mode shapes of the surrounding frequency values. The Modal Assurance Criterion (MAC) (Allemang and Brown, 1982) [32] is used for this purpose. This compares two modal shapes $\Phi_1$ and $\Phi_2$ through the following expression:

$$\mathrm{MAC}(\Phi_1, \Phi_2) = \frac{\left|\Phi_1^H \Phi_2\right|^2}{\left|\Phi_1^H \Phi_1\right\|\Phi_2^H \Phi_2\right|} \quad (6)$$

where $^H$ denotes the complex conjugate and transpose.

A MAC value greater than 80% indicates that the point still belongs to the mode "bell", even on the second singular value. The bell then represents the Transfer Function of the SDOF characterized by the peak frequency of the mode bell so that an inverse Fourier Transform leads to the Impulse Response Function (IRF) of the mode. The logarithmic decrement of the IRF gives the damping ratio and a linear regression of the zero-crossing times gives the enhanced frequency. A decision as to whether or not a peak is a structural mode can be taken by considering the extent of the mode "bell", the damping ratio and the shape. The proposed evaluation of the uncertainties on the peak position in the spectrum does not include epistemic errors, but only the uncertainties due to the windowing process in the spectral estimation as aforementioned.

Only 3 modes have been accurately determined (Fig. 5): the first longitudinal mode at 1.157±0.006 Hz, with a damping of about 0.9%, the first transverse mode at 1.217±0.006 Hz with a damping of about 1.1% and the first torsion mode at 1.45±0.01 Hz with a damping of about 0.9%. The first longitudinal mode is not pure but has a slight torsion component that is not present in the first transverse mode. Following the aforementioned decision process using MAC, the second longitudinal mode may be distinguished at 4.5±0.2 Hz and a mode that looks like the second torsion mode may be found at 5.7±0.2 Hz. In addition, the first vertical mode can be determined at 9.3±0.2 Hz. Nevertheless, because of the very poor estimate of these higher modes, only the two longitudinal modes will be considered in the following.

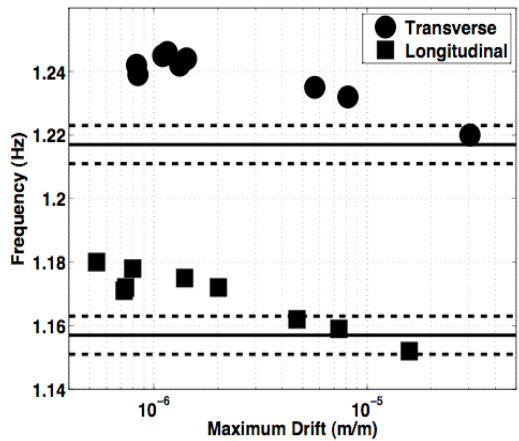

**Figure 6.** Fundamental frequencies of the building in longitudinal and transverse directions for the nine earthquakes using AR modelling and plotted as a function of the structure drift $D_m$. The solid line represents the frequency value in L- and T-direction obtained by Frequency Domain Decomposition (FDD) using ambient vibrations (+/- uncertainties shown by dashed lines)

The values of the first bending frequencies in each direction are very close to each other, meaning that the structural system has roughly the same stiffness in both directions. In reality, despite the geometric aspect ratio between longitudinal and transverse directions, Fig. 1 shows that two regular inner cores provide lateral load resistance, having roughly the same stiffness in both horizontal directions. Moreover, from Fig. 5b, it can be seen that the storeys under the prestressed slab seem to be very stiff as shown on the low deflection shape of the modes under this level.

## 6. Frequencies under earthquake recordings

To demonstrate the relevancy of the modes determined under ambient vibrations, these modes were compared to the resonance frequencies using earthquake recordings listed in Tab. 1. For this purpose, Auto-Regressive (AR) modelling of the structure was used [11]. Numerous methods could be used based on discrete time filters such as the ARX technique [33] but we decided to employ this frequency technique because its easy implementation and relevant accuracy. Each couple of base/top sensors (OGH1-OGH4, OGH2-OGH5 and OGH3-OGH6) is modelled by an AR filter obtained using the Linear Prediction method. The top motion is first deconvolved by the base motion with a water-level method [34] and the resulting spectrum is approximated by the best AR filter. A stabilisation diagram with several numbers of poles in the AR filter is used to estimate the confidence in the frequency and damping obtained for the first resonance frequency in each direction. The results are approximately the same for the three couples of sensors so that only the median value is kept for each earthquake (Fig. 6). A slight decrease (less than 2%) in the frequencies is observed with increasing drift up to $10^{-5}$. This trend seems to be linear that would mean that the frequency decreases logarithmically with respect to the drift amplitude. This elastic decrease may be due to the co-seismic aperture of micro-cracks in the concrete that temporarily decreases the stiffness of the structure and therefore the frequencies, as already observed by [10] and [11] using Californian strong-motion data collected in buildings. On the contrary, they did not observe a clear relation between damping coefficient and magnitude of the motion. Damping will not be discussed further in this paper.

The frequency during the Vallorcine earthquake is approximately 2% smaller than the frequency during the weakest ground motion, generated by the Meribel earthquake (#7, Tab. 1, Tab. 2), having the same order of acceleration as ambient vibrations. Moreover, the values obtained for this weakest motion are higher (2 to 3%) than the values obtained by the FDD method using ambient vibrations. This slight difference may be due to the system considered by FDD and AR methods: in the first case, the flexible-base building is considered including the linear soil-structure interaction that was previously observed using earthquake recordings (Fig. 4b), while in the second case, the system considered is the fixed-base building.

Although the basic assumption of white noise is required for the FDD method, it was also used to determine the structure modes during Vallorcine earthquake. The FDD is robust enough to allow this process [35]. Here again, slightly lower values of first frequencies from 1.5% to 4% are found compared to ambient vibrations (Tab. 2). In conclusion, a slight decrease (2-4%) in the first frequencies is found during the Vallorcine earthquake compared to the weakest motions (Tab. 2). This decrease has already been mentioned with reference to other buildings ([7]; [36]; [37]; [38]; [11]) but in this paper two different methods are used to quantify this decrease for weak ground motions. Care should be taken when extrapolating these results to higher drifts but this logarithmic decrease with increasing drift may be valid in the elastic domain. This means that the frequency values obtained under ambient vibrations are relevant in a building model for moderate earthquakes and that no dramatic decrease occurs between ambient vibrations and moderate earthquakes.

**Table 2.** Comparison between resonance frequencies of the structure under weak motion (ambient vibrations and Meribel earthquake) and Vallorcine earthquake using the FDD method and AR modelling.

| Resonance frequencies | FDD method | | | AR modelling | | |
|---|---|---|---|---|---|---|
| | Ambient vibrations | Vallorcine earthquake | Decrease | Meribel earthquake | Vallorcine earthquake | Decrease |
| 1st longitudinal (Hz) | 1.16 | 1.13 | 2.6% | 1.180 | 1.152 | 2.4% |
| 1st transverse (Hz) | 1.22 | 1.17 | 4.1% | 1.242 | 1.220 | 1.8% |
| 1st torsion (Hz) | 1.44 | 1.42 | 1.4% | 1.442 | 1.414 | 2.0% |

## 7. Lumped-mass models

The modal parameters obtained under ambient vibrations are unscaled [39], i.e. it is not possible to deduce the amplitude of the building motion with only modal parameters. A physical model integrating the modal parameters is therefore required. As the masses are mostly concentrated in the floors of a building, a 1D lumped-mass model was assumed for the structure. In this case, the Duhamel integral [40] gives the elastic motion $\{U(t)\}$ of each floor of the structure assuming a constant mass along the storeys $[M]$, knowing the vibration modes ($[\Phi]$ the modal shapes, $\{\omega\}$ the frequencies and $\{\xi\}$ the damping ratios) and the ground motion $U_s(t)$:

$$\{U(t)\} = [\Phi]\{y(t)\} + U_s(t) \tag{7}$$

with $\forall j \in [1, N]$, $\quad y_j(t) = \dfrac{-p_j}{\omega_j'} \int_0^t U_s''(\tau) e^{-\xi_j \omega_j (t-\tau)} \sin\left(\omega_j'(t-\tau)\right) d\tau,$ (8)

$$\omega_j'^2 = \omega_j^2\left(-\zeta_j^2\right) \text{ and } p_j = \frac{\{\Phi_j\}^T[M]\{1\}}{\{\Phi_j\}^T[M]\{\Phi_j\}} = \frac{\sum_{j=1}^{N}\Phi_{ij}}{\sum_{j=1}^{N}\Phi_{ij}^2} \text{ the participation factor of mode j.}$$

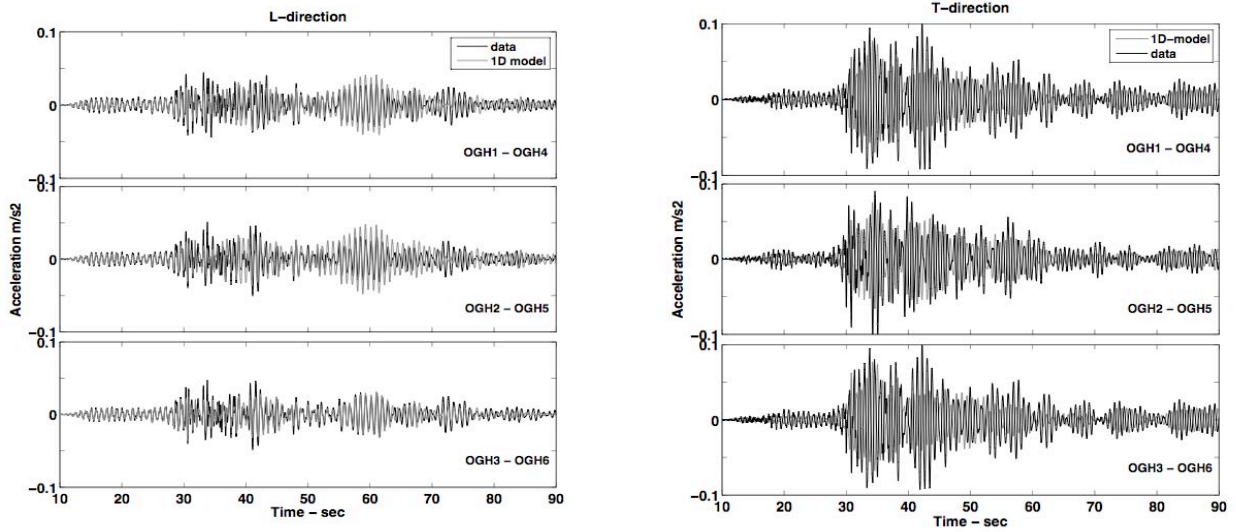

**Figure 7a.** Comparison between data (black line) and 1D lumped-mass model (gray line) of the Vallorcine (Haute-Savoie, France) $M_L$=4.9, September 8$^{th}$ 2005 earthquake at the roof of the structure in the longitudinal (left) and transverse (right) directions for each couple top/base of stations.

Only the first bending modes are considered here, the torsion mode being neglected for the sake of simplicity. This linear 1D model is then assumed so that the experimental modal shapes are averaged at each floor. The corresponding seismic motion can be computed for any deterministic (weak) earthquake scenario, considering the two horizontal motions uncoupled.

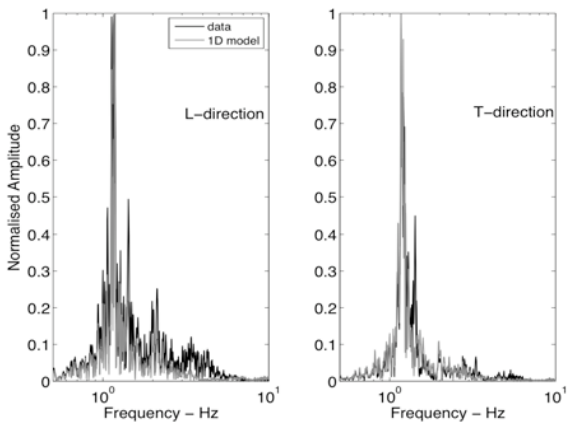

**Figure 7b.** Comparison of the Fourier spectrum between data (black line) and 1D lumped-mass model (gray line) of the Vallorcine (Haute-Savoie, France) $M_L$=4.9, September 8$^{th}$ 2005 earthquake at the roof of the structure in the longitudinal (left) and transverse (right) directions for the OGH6 station.

Considering the recording at the base floor as input (stations OGH1, OGH2 and OGH3, Fig. 2b), the synthetic motion is compared with the motion recorded at the top (Fig. 7a), considering each couple of base/top sensors, in the two horizontal directions. We observe a good fit between synthetic and data, considering the time duration, the main phases and also the amplitude of the building motion. The correlation coefficients between data and synthetic motion are 65% and 86% for OGH1-OGH4, 59% and 82% for OGH2-OGH5, and 81% and 85% for OGH3-OGH6, in the L- and T-directions, respectively. The Vallorcine recordings show significant anisotropy in the building motion: the amplitude in the transverse direction is twice the amplitude of the longitudinal direction at the top, despite a greater PGA in the longitudinal direction (Tab. 2). The difference of motion between L- and T-directions is not induced by complex behaviour of the building (the model fits the data well), but by the ground motion itself. The slight errors in the computed response are due to the 3D behaviour of the building, especially torsion and the neglect of higher modes, not included in the 1D model, as suggested on Figure 7b. This figure shows that frequency shift and approximate damping value do not influence much the response here.

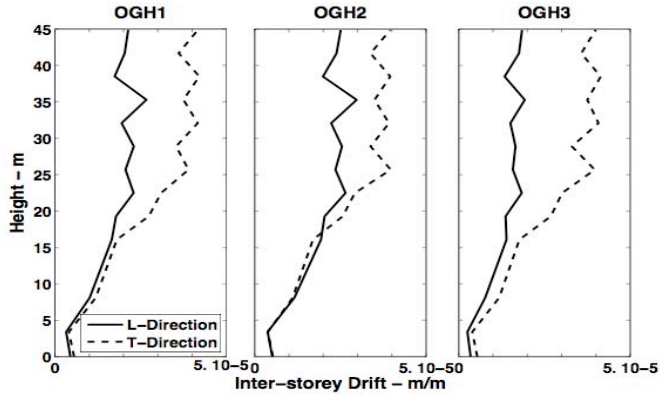

**Figure 8.** Modelling of maximum drift along the storeys of the structure during the Vallorcine (Haute-Savoie, France) $M_L$=4.9, September 8$^{th}$ 2005 earthquake using the base motion as input and the 1D lumped-mass model extracted from ambient vibrations, in the L- and T-directions.

Thanks to the lumped-mass model, it is also shown that the maximum inter-storey drift (Fig. 8) is greater in the transverse than in the longitudinal direction above the prestressed slab. Whatever the building corner, this maximum drift is approximately the same in the longitudinal direction from the 3$^{rd}$ to the 12$^{th}$ floor (2 10$^{-5}$). In the transverse direction, it is also constant from the 5$^{th}$ to the 12$^{th}$ floor at 4 10$^{-5}$, i.e. twice the longitudinal value. This maximum drift along the storeys is significantly smaller than the minimum strain able to initiate damage in the building (10$^{-4}$ for the limit tensile strain of concrete). This was confirmed by the fact that no apparent damage was observed in the Grenoble City Hall after the Vallorcine earthquake.

Slight differences between synthetic and data may also be due to the fact the 1D building model neglects SSI and torsion mode. In order to evaluate the effects of these assumptions on synthetics, the four previously mentioned parameters describing the building motion (PTA, PTV, $I_{at}$, $D_m$) are considered, together with the duration of the building motion and compared between recordings and synthetics (Fig. 9). The nine events selected in the RAP database are used for comparison (Tab. 1). The duration is defined here as the time between 5% and 95% of the Arias

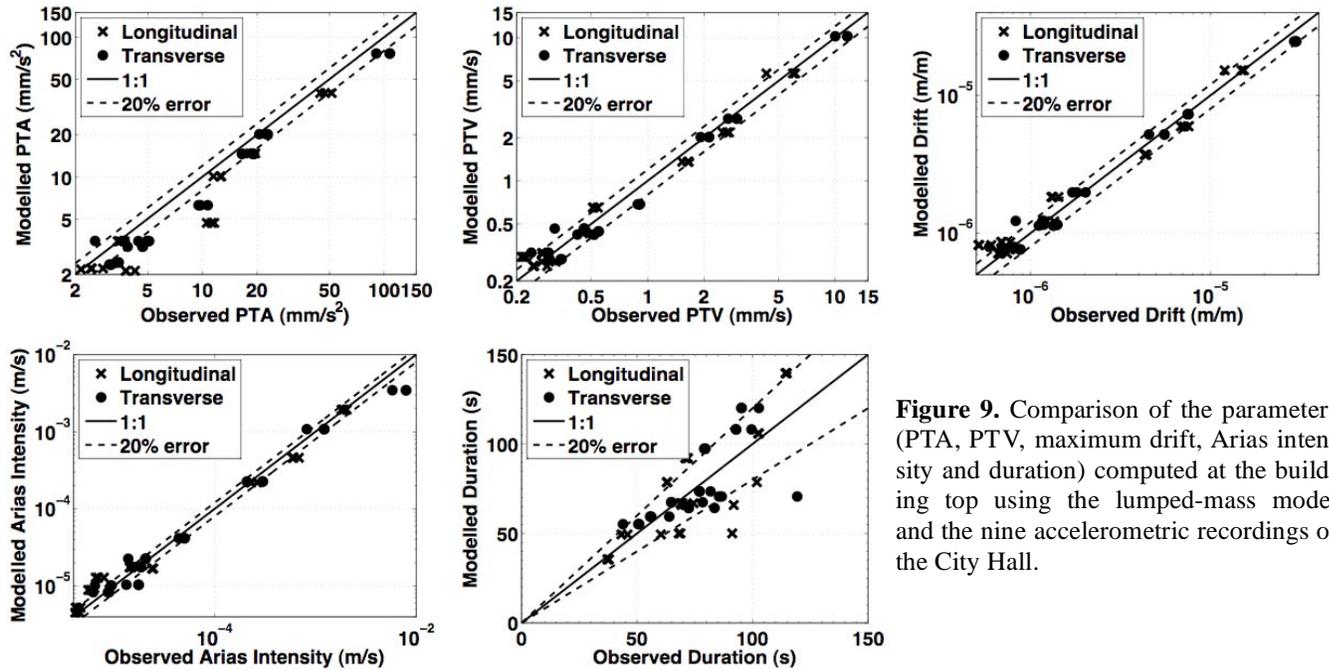

**Figure 9.** Comparison of the parameters (PTA, PTV, maximum drift, Arias intensity and duration) computed at the building top using the lumped-mass model and the nine accelerometric recordings of the City Hall.

Intensity [41]. The accelerations often tend to be underestimated, especially for the weakest earthquakes, since torsion is not taken into account in the model (Fig 9). Conversely, the duration is sometimes overestimated in the model possibly as a result of the damping ratio that may be higher and therefore may decrease the resonance duration. Most of the errors on PTV, $D_m$ and duration are less than 20%. The Arias Intensity at the top of the building is well reproduced except for the smallest earthquakes. The overall results are nevertheless satisfactory and they validate the simple 1D lumped-mass model obtained from ambient vibrations.

## 8. Multifiber beam model

A detailed finite element model is presented hereafter able to reproduce numerically the modal and the non linear behaviour of the Grenoble City Hall. Spatial discretisation is done using multifiber Timoshenko beam elements [42] and constitutive laws based on damage mechanics [43] for concrete and plasticity for steel [44]. Perfect bond is considered. The slabs are modelled with Kirchhoff plate elements assuming a linear elastic behaviour. More specifically:

- The total number of the elements in the finite element mesh is 18928.
- The finite element code used is Cast3M.
- Concrete parameters: Young's modulus 32GPa, Poisson's coefficient 0.2, traction limit $f_t$ = 3MPa, compression limit $f_{c28}$ = 30 MPa, mass density 2400 Kg/m$^3$.
- Steel: Young's modulus 200GPa, Poisson's coefficient 0.3, yield stress 400MPa, yield strain 0.03, ultimate stress 460 MPa, ultimate yield strain 0.09, mass density 7800 Kg/m$^3$.
- Total mass of the structure: 9540 10$^3$ Kg.

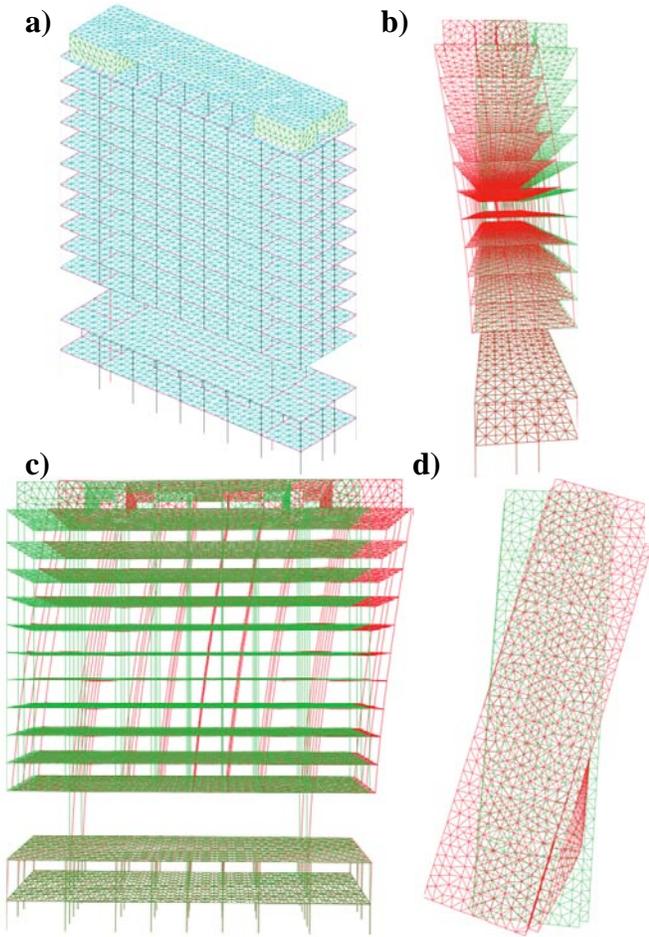

**Figure 10.** Finite element mesh of the Grenoble City Hall using multifiber beam elements (a), numerical shapes of the first bending mode in the transverse direction (b), in the longitudinal direction (c) and for the torsion mode (d).

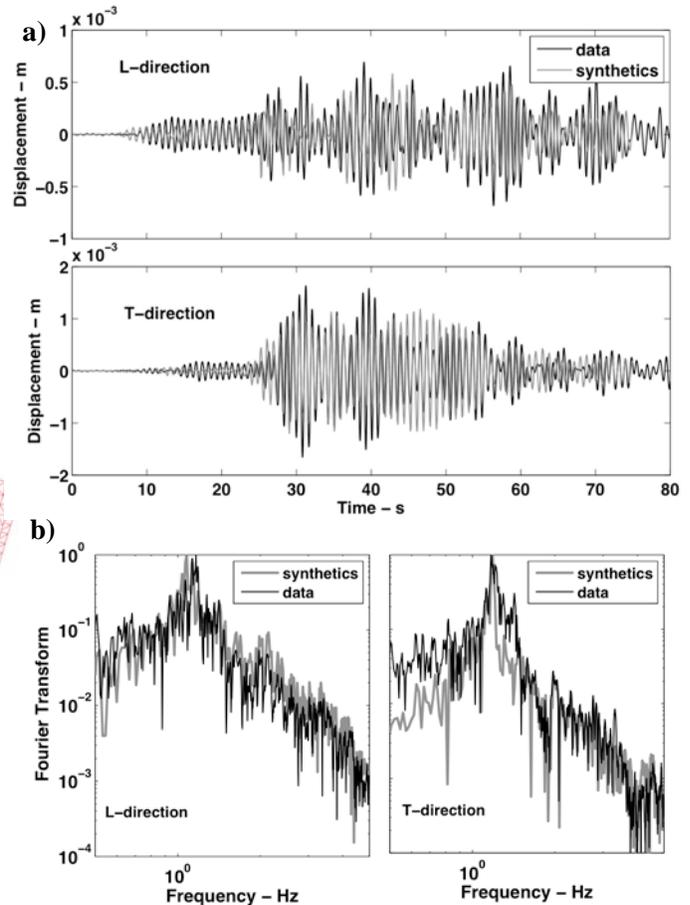

**Figure 11.** Comparison of the displacements recorded by the accelerometric building array (black line) and computed by the multifiber beam model (grey line) of the Vallorcine (Haute-Savoie, France) $M_L$=4.9, September 8$^{th}$ 2005 earthquake. Displacements are considered at the roof of the structure in the longitudinal L and transverse T directions. a) Time-history of the motion at the OGH4 corner of the building. b) Fast Fourier transform of the time-history, normalised by the maximal amplitude of each spectra.

Figure 10 shows the complete finite element mesh and the first three computed modal shapes. The numerical values for the first three frequencies are 1.10 Hz, 1.18 Hz and 1.43 Hz for the longitudinal, transverse and rotational modes, respectively. One can see that they are very similar to the ones found using the FDD method or the AR modelling (Table 2). In order to validate the multifiber model, Figure 11 shows the comparison of the numerical displacements and the corresponding Fast Fourier transform with the ones recorded from the accelerometric building array for the Vallorcine earthquake. The correlation coefficients between time histories of data and 3D modelling are 98% and 95% in the L- and T-directions, respectively. Finally, and for the same earthquake, Figure 12 shows the comparison of the maximum inter-story drift along the height of the structure using the base motion as input, the 1D lumped-mass model (black line) and the multifiber beam model (grey line). Results are again similar proving the effectiveness of the 1D model extracted from ambient vibrations for the case of moderate earthquakes.

## 9. Conclusions

16/20

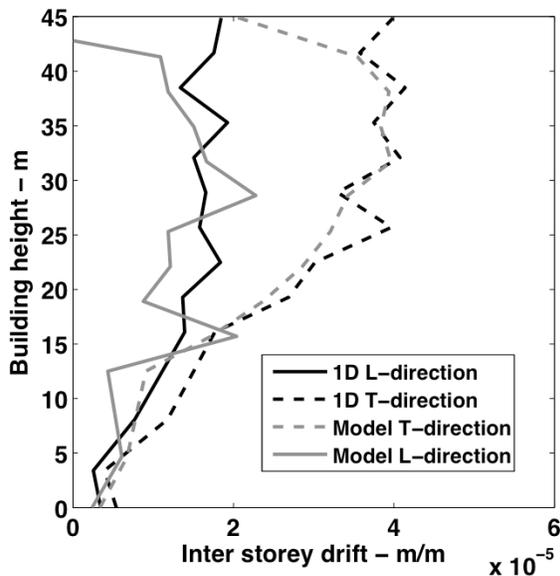

vibrations (black line) and the multifiber beam model (gray line) in the longitudinal L (continuous) and transverse T (dashed) directions.

This paper shows how the dynamic response of existing buildings in the elastic domain is obtained from ambient vibrations. Thanks to the past development of new Operational Modal Analysis methods based on ambient vibrations, the modal response of buildings can be understood and obtained in order to predict building behaviour under weak motion. The study is focused on the Grenoble City Hall building that has the advantage of being permanently monitored. The seismic recordings, supplemented with full-scale ambient vibration measurements have enabled a better understanding of the dynamic behaviour of the structure. This behaviour is largely dominated by the first bending mode in each direction, including nevertheless slight torsion components. During recorded earthquakes, the frequencies of the structure decreased by 3% with respect to the ambient vibration values. The decrease in frequency follows a logarithmic decay with respect to the drift of the structure. This decrease is sufficiently small to consider that the modal properties obtained from ambient vibrations are relevant in a wide range of amplitudes, while the building stays undamaged. This was also shown by Dunand et al. [11] using Californian data.

Assuming a 1D lumped and constant mass model, the experimental modal parameters were used to reproduce the motion of the building for moderate earthquakes, without any hypothesis on the structural design and materials. Such building motion parameters as acceleration or velocity amplitude, duration, drift and energy are reproduced relatively well with this simple model. Therefore, the response of a structure to moderate earthquakes can be easily predicted as soon as the intrinsic behaviour of the building under ambient vibrations has been accurately determined using experimental techniques. This model can be used to calculate the inter-storey drift for any weak motion and a subsequent preliminary assessment of the integrity of the building following the integrity threshold concept developed by Boutin et al. [45] who consider the damaged/undamaged limit as the end of the linear behaviour of the structure.

The inter-storey drift results from the 1D model are also compared to the ones coming from a 3D model using multifiber beams. The drift is well reproduced by the 1D-lumped mass model that uses only the modal model extracted from ambient vibrations. The 3D model on the other hand needs a complete description of the structure. Therefore, the modal model is also useful to calibrate more sophisticated models in the elastic regime, so that they can be used to assess more precise damage parameters and afterwards to explore the anelastic behaviour of buildings under strong motion.

For the case of a large set of buildings that can be affected by earthquakes, numerical modelling has prohibitive cost and should be limited to strategic buildings. For usual buildings, modal parameters extracted by enhanced modal analysis methods and using ambient vibrations could be an alternative to costly methods when coupled to the interstory drift concept for seismic vulner-ability assessment. This is particularly the case for moderate seismic regions where the seismic retrofitting investment is rather limited with however past damaging earthquakes.

**Acknowledgements**
This work is supported by the Rhone-Alps regional authorities through the Thématiques Prioritaires program (project VULNERALP) and by the French Research National Agency (ANR) through the RGCU program (project ARIVSE ANR-06-PRGCU-007-01, http://arvise.grenoble-inp.fr/index.jsp). The Grenoble strong-motion network is operated by the *Laboratoire de Géophysique Interne et Tectonophysique* (LGIT) for the French Accelerometric Network (RAP). The authors are very grateful to E. Chaljub, M. Langlais and S. Hatton for operating the stations and C. Péquegnat for managing the data centre and providing the data, without which this study would not have been possible.